\documentclass[smallcondensed,final]{svjour3}

\usepackage{mathbbol}
\usepackage[numbers,sort&compress]{natbib}
\usepackage{dsfont,citesort}
\usepackage[dvipsnames]{xcolor}
\usepackage{graphicx}
\usepackage{bm}
\usepackage{amsmath,amsfonts}
\usepackage{subfigure}
\usepackage{amssymb}
\usepackage{stmaryrd}
\usepackage{stackengine}
\usepackage{ifthen}
\usepackage{mathrsfs}

\newcommand{\sref}[1]{Section \ref{#1}}

\newcommand{\avg}[1]{\left \langle #1 \right \rangle }

\newcommand{\f}[2]{
		\mathchoice%
			{\dfrac{#1}{#2}}
	    	{\dfrac{#1}{#2}}
			{\frac{#1}{#2}}
			{\frac{#1}{#2}}
}
\newcommand{\dd}{\mathrm{d}}
\newcommand{\e}[1]{_{\text{#1}}}

\newcommand{\ddf}[3][]{\ifthenelse{\equal{#1}{}}{\ensuremath{\f{\dd#2}{\dd#3}}}
{\ensuremath{\f{\dd^{#1}#2}{\dd{#3}^{#1}}}}}
\newcommand{\vect}[2][]{{\bm{#2}_{\mathrm{#1}}}}

\newcommand{\norm}[1]{\ensuremath{ \left| #1 \right|}}

\newcommand{\R}{\mathbb{R}}

\renewcommand{\leq}{\ensuremath{\leqslant}}
\renewcommand{\geq}{\ensuremath{\geqslant}}

\newcommand{\ie}{\textit{i}.\textit{e}.}
\newcommand{\eg}{\textit{e}.\textit{g}.}

\newcommand{\Op}{\Theta}
\newcommand{\Ss}{\mathscr{S}}

\journalname{J. Stat. Phys.}

\begin{document}

\title{Isometric Uncertainty Relations}

\author{Hadrien Vroylandt   \and
        Karel Proesmans     \and
        Todd R. Gingrich
}

\institute{Hadrien Vroylandt \at
              Department of Chemistry, Northwestern University, 2145 North Sheridan Road, Evanston, Illinois, USA.\\
              hadrien.vroyland@northwestern.edu
          \and
          Karel Proesmans  \at
          Simon Fraser University, 8888 University Drive, Burnaby, British Columbia, Canada\\
          Hasselt University,  Martelarenlaan 42, Hasselt, Belgium\\
          karel\_proesmans@sfu.ca
          \and
          Todd R. Gingrich \at
          Department of Chemistry, Northwestern University, 2145 North Sheridan Road, Evanston, Illinois, USA.\\
          todd.gingrich@northwestern.edu
}

\maketitle

\begin{abstract}
We generalize the link between fluctuation theorems and thermodynamic uncertainty relations by deriving a bound on the variance of fluxes that satisfy an isometric fluctuation theorem. 
The resulting bound, which depends on the system's dimension $d$, naturally interpolates between two known bounds. 
The bound derived from the entropy production fluctuation theorem is recovered for $d=1$, and the original entropy production thermodynamic uncertainty relation is obtained in the $d \to \infty$ limit.
We show that our result can be generalized to order parameters in equilibrium systems, and we illustrate the results on a Heisenberg spin chain.

\keywords{Isometric fluctuation theorem, nonequilibrium steady state, thermodynamic uncertainty relation, broken symmetry.}
\PACS{05.70.Ln -- 05.40.-a -- 02.50.-r}
\end{abstract}
\section{Introduction}
\label{sec:Introduction}
Current generation is central to nonequilibrium processes much like fluctuations are central to microscopic dynamics.
Consequently, when studying microscopic systems away from equilibrium, intense interest has focused on the overlap between the two: fluctuations in currents.
Those current fluctuations have been studied in explicit models to gain insight into the reliability of molecular motors \cite{seifert2012stochastic,brown2019theory}, the behavior of exclusion processes \cite{derrida2007non}, and fluxes in macroscopic systems \cite{bertini2015macroscopic}.
Recently, it has been demonstrated that under fairly broad conditions, and without reliance on a particular model system, current fluctuations are constrained by thermodynamic considerations.
The first such demonstration, applying to Markovian jump dynamics on finite state spaces, was a so-called Thermodynamic Uncertainty Relation (TUR) which states that the relative size of current fluctuations compared to their mean is constrained by the average rate that entropy is produced in the nonequilibrium steady state~\cite{Barato2015}.
Initial derivations of this TUR, hereafter referred to as the original TUR, relied heavily on methods of large deviation theory to explicitly bound the size of current fluctuations using the probability of jump trajectories~\cite{Gingrich2016, Gingrich2017}.
Subsequent analyses used information geometry~\cite{Dechant2018FRI, Dechant2018} and the entropy production fluctuation theorem (FT)~\cite{van2019uncertainty, Timpanaro2019Thermodynamic,potts2019thermodynamic} to derive related results.
One appeal of those alternate perspectives is that the resulting bounds on current fluctuations, though typically weaker, can apply more broadly than the original TUR~\cite{Horowitz2019}.
For example, uncertainty relations derived from the FT can be applied to classical systems with discrete timesteps, to time-dependent driving, or even to quantum dynamics, all situations that can violate the original TUR~\cite{shiraishi2017finite,proesmans2017discrete,barato2016cost,Agarwalla2018assessing, Liu2019thermodynamic}.

The uncertainty relations derived from the FT can be seen as consequences of a symmetry, or more precisely a broken symmetry.
In the absence of dissipation, trajectories and their time-reversals occur with equal probability.
Dissipation skews that balance---the greater the dissipation, the greater the broken time-reversal symmetry, which directly impacts the distribution of currents.
Consider, for example, a particular trajectory which accumulates current \(J\) and its time-reversal which must accumulate current \(-J\).
The relative probability of observing the two trajectories is regulated by the dissipation, but that relative probability also impacts the degree of uncertainty in measurements of current.
Hence, there must exist some inequalities which bound that uncertainty by a function of the dissipation.
Recently, Timpanaro et al.\ presented the particular function of dissipation that gives the tightest possible TUR derivable solely from broken dynamical symmetry of the FT~\cite{Timpanaro2019Thermodynamic}.
That result is necessarily weaker than the original TUR, which implicitly respects the FT symmetry but which also utilizes additional statistical properties of Markovian jumps.
Despite producing weaker bounds, we find it stimulating to investigate what limitations on fluctuations can arise from symmetry arguments alone.
In this paper we consider extra broken symmetries beyond time reversal and analyze how those additional symmetry relations strengthen uncertainty bounds.

The motivating broken symmetry is that of the isometric fluctuation theorem\cite{Hurtado2011Symmetries,Hurtado2014Thermodynamics,VillavicencioSanchez2014Fluctuation,PerezEspigares2015Spatial}, which we briefly motivate for completeness.
Imagine observing heat transfer through a semi-infinite slab that is sandwiched between hot and cold thermal reservoirs.
On average, energy will transfer from hot to cold with no flow parallel to the slab-reservoir surfaces, but if one considers fluctuations away from this average behavior, there is a distribution over the possible vectorial heat currents \(\vec{J}\).
Much like forward and reversed trajectories have different probability in the case of broken time-reversal symmetry, all possible \(\vec{J}\) with the same magnitude do not have the same probability.
In this standard example, the isometric FT requires that various realizations of \(\vec{J}\) with identical magnitudes have relative likelihoods which are simply expressed in terms of a global symmetry-breaking field, in this case the temperature gradient between the reservoirs\cite{Hurtado2011Symmetries,Hurtado2014Thermodynamics}.
Our work here shows how the isometric FT implies a stronger TUR than could be obtained solely on the basis of the FT.
The FT-based TURs depend only on the dissipation, and we demonstrate that the stronger isometric TUR likewise depends only on physically observable quantities related to the breaking of symmetry.
We should be clear that incorporating additional broken symmetries still does not constrain fluctuations as strongly as the original TUR, so any benefit of this work appears greatest when that original TUR breaks down, \eg, discrete time, time-dependent or quantum dynamics.

By framing the arguments in terms of generic broken symmetries, our work translates naturally to uncertainty relations on fluctuations which are not dynamic in nature, \ie, not only fluctuations in currents.
As an example, we illustrate the impact of broken symmetries on fluctuations in the classical equilibrium statistical mechanics of spin systems.
By systematically increasing the dimensionality of the spin space, we demonstrate how additional broken-symmetries tighten the uncertainty relations.

The structure of the paper is as follows.
We begin with a review of broken symmetries, the isometric FT, and the Von Mises-Fisher distribution in Sec.~\ref{sec:Setup}.
Given all of an order parameter's broken symmetries, the Von Mises-Fisher distribution is the distribution which maximizes that order parameter's uncertainty.
Using this fact, we derive an isometric TUR from the isometric FT in Sec.~\ref{sec:Derivation}.
Section~\ref{sec:equilibrium} maps the isometric TUR for dynamical fluctuations into uncertainty relations for order parameters of classical equilibrium statistical mechanical spin systems.
In that context, it is simple to tune the number of broken symmetries by changing the dimensionality of the spin space, thereby allowing us to systematically reveal that greater symmetry tightens the thermodynamics bounds.
Finally, in Sec.~\ref{sec:Discussion} we close with a discussion.

\section{Setup}
\label{sec:Setup}

\subsection{Broken time-reversal symmetry and the fluctuation theorem}
\label{sec:brokenTR}

Consider a Markovian system in a $d$-dimensional continuous space.
At time $t$, the state of the system is $\vect{X}_t \in \R^d$, and we denote a trajectory by $\{\vect{X}_t\}$.
The time-reversal operator, ${\rm TR}$, acts on the trajectory to return the time-reversed trajectory that we denote by $\{\vect{X}'_t\} = {\rm TR}(\{\vect{X}_t\})$.
For an equilibrium dynamics, there is a time-reversal symmetry such that the path probability of the two trajectories are equal: $P_0(\{\vect{X}_t\}) = P_0(\{\vect{X}'_t\})$.
For nonequilibrium dynamics, that symmetry is broken by some thermodynamic forces $\vect{\mathcal{F}}$ which induce currents $\vect{J}$ conjugate to the forces.
The dot product between those forces and currents is the Radon-Nikodym derivative of the path probability with respect to the path probability of the time-reversed path~\cite{Esposito2010Three}:
\begin{equation}
\vect{\mathcal{F}} \cdot \vect{J} =  \ln \frac{\dd P_{\vect{\mathcal{F}}}(\{\vect{X}_t\})}{\dd P_{\vect{\mathcal{F}}}(\{\vect{X}'_t\})}.
\label{eq:RN1}
\end{equation}
The current is an observable function of the trajectory, $\vect{J}(\{\vect{X}_t\})$ that is inverted upon time-reversal of the trajectory.
With suitable marginalization, one can pass from the path probabilities to the probability $P(\vect{J})$ of observing current $\vect{J}$.
Due to Eq.~\eqref{eq:RN1}, this distribution of currents must satisfy a FT:
\begin{align}
\nonumber P(\vect{J}) &= \int \dd P_{\vect{\mathcal{F}}}(\{\vect{X}_t\} ) \delta(\vect{J}(\{\vect{X}_t\} ) - \vect{J})\\
  \nonumber &= \int \dd P_{\vect{\mathcal{F}}}(\{\vect{X}'_t\} )\f{\dd P_{\vect{\mathcal{F}}}(\{\vect{X}_t\} )}{\dd P_{\vect{\mathcal{F}}}(\{\vect{X}'_t\} )}  \delta(\vect{J}(\{\vect{X}_t\} ) - \vect{J})\\
  \nonumber &= e^{\vect{\mathcal{F}}\cdot \vect{J}}  \int \dd P_{\vect{\mathcal{F}}}(\{\vect{X'}_t\} ) \delta(\vect{J}(\{\vect{X'}_t\} ) + \vect{J})\\
&= P(-\vect{J}) e^{\vect{\mathcal{F}}\cdot \vect{J}}.
\label{eq:FT}
\end{align}
Viewed in this way, the dot product between thermodynamic forces and conjugate currents is the measure of the time-reversal symmetry breaking, but this dot product also has a thermodynamic meaning: the dissipation $\Sigma = \vect{\mathcal{F}} \cdot \vect{J}$.
That connection between time-reversal symmetry breaking and thermodynamics is the core feature of the FT, and it is possible to carry out completely analogous mathematical steps based upon other broken symmetries.

\subsection{Broken rotational symmetry and the isometric fluctuation theorem}
\label{sec:brokenR}
Rather than focusing on the action of time reversal on the trajectory's probability, one may consider the impact of a rotation $\vect{R}$ of the $d-$dimensional state space.
As in the preceding analysis, we write the transformed trajectory as $\{\vect{X'}_t\}=\bm{R}(\{\vect{X}_t\})$ such that for all times $t$, $\vect{X'}_t = \bm{R}\vect{X}_t$, \ie, the trajectory under the action of the rotation is $\{\vec{X'}_t\}$.
For an isometric (symmetric) system with path probability $ P_{\vect{0}}(\{\vect{X}_t\})$, this rotation should not have an effect; the probability to observe a trajectory is the same as the probability to observe the transformed trajectory $P_{\vect{0}}(\{\vect{X}_t\}) = P_{\vect{0}}(\{\vect{X'}_t\}) $.
In the presence of an external field $\vect{F}$, this rotational symmetry is broken, but a relation emerges between the probability of an observable and the probability of the transformed observable\cite{Hurtado2011Symmetries,VillavicencioSanchez2014Fluctuation,PerezEspigares2015Spatial}. 
Following the example of time-reversal symmetry, the current-like observable, which we will denote as  $\vect{\Op}(\{\vect{X}_t\})$, is the one conjugate to $\vect{F}$. 
In analogy with \eqref{eq:RN1}, what we mean by conjugacy is that $\vect{\Op}$ satisfies
\begin{equation}
  \label{eq:orderParameter}
  \f{1}{2} \vect{F}\cdot \left( \vect{\Op} (\{\vect{X}_t\})-\bm{R}\vect{\Op} (\{\vect{X}_t\})\right) =  \ln \f{\dd P_{\vect{F}}(\{\vect{X}_t\}) }{\dd P_{\vect{F}}(\{\vect{X}'_t\})  }.
\end{equation}
Like $\vect{J}$ above, $\vect{\Op}$ can be computed as a function of the trajectory $\{\vect{X}_t\}$ and is thus an order parameter of the trajectory which characterizes the breaking of the rotational symmetry by the external field.
Under a rotation of the trajectory, the order parameter $\vect{\Op}$ will also be transformed as $\vect{\Op}' = {\bm R} \vect{\Op}$.
The analog of Eq.~\ref{eq:FT} yields the isometric fluctuation theorem that constrains the fluctuations of  $\vect{\Op}$ by
\begin{equation}
  \label{eq:isometricFT}
  \f{P(\vect{\Op})}{P(\vect{\Op'})} = e^{\frac{1}{2} \vect{F}\cdot (\vect{\Op}-\vect{\Op'})} 
\end{equation}
where $\vect{\Op}$ and $\vect{\Op'}$ are related by a rotation.

Because Eqs.~\eqref{eq:FT} and~\eqref{eq:isometricFT} share the same structure, we should highlight the distinctions.
Principally, the FT of Eq.~\eqref{eq:FT} compares the probability of only two possible values of an observable: $\vect{J}$ and $-\vect{J}$ which are related by a discrete transformation (time-reversal).
The isometric FT, by contrast, expresses a family of comparisons between values $\vect{\Op}$ and $\vect{\Op'}$ for all continuous rotations ${\bm R}$.
However, among all the rotations, we always have the discrete parity operation\footnote{The $\mathbb{Z}_2$ group is always a subgroup of rotation groups.} that maps $\vect{\Op}$ to $-\vect{\Op}$.
Applying this parity operation inverts spatial currents in the same manner as the time-reversal operation.
We deduce that $\vect{\Op}$ must therefore be like the current, though it can actually differ from $\vect{J}$ by surface terms\footnote{These surface terms will depends of the chosen dynamics. See Ref.~\cite{PerezEspigares2015Spatial} for more details. In particular, $\vect{\Op}$ will not include currents generated by time-periodic driving.}.
In the long-time limit, those surface terms decay away to leave an asymptotic isometric FT in terms of currents:
\begin{equation}
\lim_{t \to \infty} \frac{1}{t} \ln \frac{P(\vect{J})}{P(\vect{J}')} = \frac{1}{2} \vect{F} \cdot (\vect{J} - \vect{J}'),
\end{equation}
with $\vect{J}$ and $\vect{J}'$ related by rotation~\cite{Hurtado2011Symmetries}.
We have retained the surface terms in our presentation so that the isometric FT, Eq.~\eqref{eq:isometricFT}, holds at all times.

\subsection{Angular distributions}

A strong consequence of the isometric FT is that the angular part of the distribution $P(\vect{\Op})$ can be expressed in terms of the Von Mises-Fisher probability distribution,
\begin{equation}
  \label{eq:fisherVonMises}
  f\e{vMF}\left(\vect{x} ; r{\vect{e}}_0 \right) = \f{ e^{r{\vect{e}}_0\cdot \vect{x}}}{Z_d(r)} \quad \text{for} \quad \vect{x} \in \mathbb{S}_d.
\end{equation}
The random variable $\vect{x}$ is confined to the $(d-1)$-dimensional unit sphere $\mathbb{S}_d$, with the distribution parameterized by a radius $r$ and a preferred direction $\vect{e}_0$.
Notably, the normalization factor $Z_d(r)$ can be explicitly integrated in terms of $I_\nu(r)$, the modified Bessel function of the first kind of order $\nu$:
\begin{equation}
  \label{eq:normalisationCsteFvM}
  Z_d(r) =\int_{\vect{x} \in \mathbb{S}_d} \dd \vect{x}  e^{r{\vect{e}}_0\cdot \vect{x}}  = \f{(2\pi)^{d/2}I_{d/2-1}(r)}{r^{d/2-1}}.
\end{equation}
We decompose $\vect{\Op}$ into a product of its norm $\norm{\vect{\Op}}$ and its direction $\vect{x} \equiv \vect{\Op} / \norm{\vect{\Op}}$ to express $P(\vect{\Op})$ in terms of the Von Mises-Fisher distribution:
\begin{align}
\nonumber  P(\vect{\Op}) &=   \f{P(\norm{\vect{\Op}}\vect{x})}{P(\norm{\vect{\Op}}\vect{e}_{\vect{F}})}P(\norm{\vect{\Op}}\vect{e}_{\vect{F}})  \\
                                          &=  \nonumber P(\norm{\vect{\Op}}\vect{e}_{\vect{F}})e^{\frac{1}{2}\norm{\vect{\Op}}\norm{\vect{F}}\vect{e}_{\vect{F}}\cdot(\vect{x}-\vect{e}_{\vect{F}})}\\
                                          &= \nonumber P(\norm{\vect{\Op}}\vect{e}_{\vect{F}})e^{-\frac{1}{2}\norm{\vect{\Op}}\norm{\vect{F}}\vect{e}_{\vect{F}}\cdot\vect{e}_{\vect{F}}}Z_d\left(\f{\norm{\vect{\Op}}\norm{\vect{F}}}{2}\right)\f{e^{\frac{1}{2}\norm{\vect{\Op}}\norm{\vect{F}}\vect{e}_{\vect{F}}\cdot\vect{x}}}{Z_d\left(\f{\norm{\vect{\Op}}\norm{\vect{F}}}{2}\right)}\\
                                    &= Q(\norm{\vect{\Op}}\norm{\vect{F}})  f\e{vMF}\left(\vect{x} ; \frac{1}{2}\norm{\vect{\Op}}\norm{\vect{F}}{\vect{e}}_{\vect{F}} \right),  \label{eq:sumvMF}
\end{align}
where $\vect{e}_{\vect{F}} \equiv \vect{F}/\norm{\vect{F}}$ is a unit vector in the direction of the symmetry-breaking field.
Note that $Q(r)= P\left(\f{r}{\norm{\vect{F}}}\vect{e}_{\vect{F}}\right)e^{-\frac{r}{2}}Z_d(\frac{r}{2})$ is a normalized probability distribution over the positive reals\footnote{$\int_0^{+\infty} \dd r \, P\left(\f{r}{\norm{\vect{F}}}\vect{e}_{\vect{F}}\right)e^{-\f{r}{2}}\int_{\vect{x} \in \mathbb{S}_d} \dd \vect{x}  e^{\f{r}{2}{\vect{e}}_{\vect{F}}\cdot \vect{x}} =\int_0^{+\infty} \dd r\int_{\vect{x} \in \mathbb{S}_d} \dd \vect{x}P\left(\f{r}{\norm{\vect{F}}}\vect{x}\right) = 1$}.
The distribution for $\vect{\Op}$ therefore decomposes into the product of a known high-dimensional angular distribution $f\e{vMF}$ and the single-variable radial distribution $Q(r)$.
In fact, we have employed this decomposition precisely because it highlights that the symmetry constraints affect the angular distribution but do not restrict $Q(r)$.
As such, $Q(r)$ can in principle be any normalized radial distribution, even the delta function employed in the next section.

\section{Derivation of Isometric Uncertainty Relations}
\label{sec:Derivation}

An isometric uncertainty relation for $\vect{\Op}$ requires that we compute the first and second moments of that vectorial order parameter. 
However, as in the original TUR, it is simpler to study fluctuations in a scalar observable.
In that context, it is common to project high-dimensional currents onto a single scalar current, and the TURs tend to yield particularly tight inequalities when the dissipation, $\Sigma = \vect{\mathcal{F}} \cdot \vect{J}$, is chosen as the scalar current.
The analog for the isometric case is to study fluctuations in the scalar\footnote{This scalar is the analog as $\Ss$ equals the dissipation $\Sigma$ in the long-time limit.} $\Ss = \vect{F} \cdot \Op$, whose moments are simply expressed in terms of the moments of the von Mises-Fisher distribution via the change of variables
\begin{align}
\nonumber \avg{\Ss^n} &= \int \dd \vect{\Op}\,(\vect{F}\cdot\vect{\Op})^n Q(\norm{\vect{\Op}}\norm{\vect{F}})  f\e{vMF}\left(\f{\vect{\Op}}{\norm{\vect{\Op}}} ; \f{\norm{\vect{\Op}}\norm{\vect{F}}}{2}{\vect{e}}_{\vect{F}} \right)\\
  &=\int \dd r\,Q(r) r^n \int _{\vect{x} \in \mathbb{S}_d}\dd \vect{x} ({\vect{e}}_{\vect{F}}\cdot \vect{x})^n f\e{vMF}\left(\vect{x} ; \f{r}{2}{\vect{e}}_{\vect{F}} \right).\label{eq:momentOp}
\end{align}
We highlight that by choosing to study fluctuations in $\Ss$, we are drawing a parallel with TURs for entropy production fluctuations but not for fluctuations in generalized currents.

From Eq.~\eqref{eq:momentOp}, the first and second moment of $\vect{F}\cdot\vect{\Op}$ can be expressed as the radial averages
\begin{equation}
  \label{eq:meanValue}
  \avg{\Ss} = 2\int \dd r \, Q(r) e_d\left(\f{r}{2}\right),
\end{equation}
and
\begin{equation}
  \label{eq:secondmoment}
  \avg{\Ss^2} = 4\int \dd r \, Q(r) c_d\left(\f{r}{2}\right).
\end{equation}
where the functions $e_d(r)$ and $c_d(r)$ are angular averages taken over the sphere of radius $r$, which are computed by integrating the Von-Mises-Fisher distribution.
Specifically,
\begin{equation}
  \label{eq:er}
  e_d(r) = r \int _{\vect{x} \in \mathbb{S}_d} \dd \vect{x} {\vect{e}}_0\cdot \vect{x}  f\e{vMF}\left(\vect{x} ; r{\vect{e}}_{0} \right) = r \f{\ddf{Z_d}{r} (r)}{Z_d(r)} = r\f{I_{d/2}(r)}{ I_{d/2-1}(r)}
\end{equation}
and
\begin{equation}
  \label{eq:cr}
  c_d(r)= r^2 \int _{\vect{x} \in \mathbb{S}_d} \dd \vect{x} ({\vect{e}}_0\cdot \vect{x})^2  f\e{vMF}\left(\vect{x} ; r{\vect{e}}_{0} \right)= r^2 \f{\ddf[2]{Z_d}{r}(r)}{Z_d(r)} = r^2 -(d-1)e_d(r).
\end{equation}

Note that if one alters the stochastic dynamics, $Q(r)$ will respond to the change but $e_d(r)$ and $c_d(r)$ are completely fixed by the symmetry properties.
As such, we can deduce the minimum possible variance (at fixed $\avg{\Ss}$) by inserting in place of $Q(r)$ the narrowest possible radial distribution that fulfills the constraint on the mean.
This extremal distribution is given by
\begin{equation}
\tilde{Q}(r) = \delta(r - r^*),
\label{eq:Qdelta}
\end{equation}
with $r^*$ chosen such that $2\int \tilde{Q}(r) e_d(r) = \avg{\Ss}$, a choice directly inspired by Ref.~\cite{Timpanaro2019Thermodynamic}.
As $e_d(r)$ is an invertible function (see App.~\ref{app:convexity}), we can obtain the extremal radius  $r^*$ by $r^* = e_d^{-1}(\avg{\Ss}/2)$.

It remains to prove that $\tilde{Q}(r)$ indeed yields the minimum variance.
To do so we show that the second moment for an arbitrary $Q(r)$ is bounded from below by the second moment of $\tilde{Q}(r)$:
\begin{align}
  \label{eq:proof_bounds_1}
  \avg{\Ss^2} & = 4\int \dd r \, Q(r) c_d\left(e_d^{-1}\left(e_d\left(\f{r}{2}\right)\right)\right) \nonumber\\
  &\geq 4 c_d\left(e_d^{-1}\left(\int \dd r \, Q(r) e_d\left(\f{r}{2}\right)\right) \right) = 4 c_d\left(e_d^{-1}\left( \f{\avg{\Ss}}{2} \right) \right) = 4\int \dd r \tilde{Q}(r) c_d(r). 
\end{align}
The inequality step of this algebra results from a convexity argument since $c_d(e_d^{-1}(u))$ is a convex function of $u$ (see App.~\ref{app:convexity}).
Having bounded the second moment, we derive a $d-$dimensional lower bound on the variance:
\begin{equation}
\text{Var}(\Ss) \geq V_d(\avg{\Ss}).
\label{eq:main}
\end{equation}
A straightforward calculation gives this bound in terms of the inverse function $e_d^{-1}$ (which does not have a simple analytic form):
\begin{align}
 V_d(\avg{\Ss}) &= \avg{\Ss}^2 \left(\f{c_d\left(e_d^{-1}\left(\f{\avg{\Ss}}{2}\right) \right)}{e_d\left(e_d^{-1}\left( \f{\avg{\Ss}}{2} \right) \right)^2} -1\right) \\
    &=4e_d^{-1}\left(\frac{\avg{\Ss}}{2}\right)^2-2(d-1)\avg{\Ss}-\avg{\Ss}^2.
\end{align}

The final equality of Eq.~\eqref{eq:proof_bounds_1} emphasizes that the lower bound is associated with our extremal distribution~\eqref{eq:Qdelta}.
This bound will therefore be sharp when the dynamics yields a peaked $Q(r)$ which resembles the delta function $\tilde{Q}(r)$.
In most cases that will not be the case, but in the symmetric case where $\vect{F} = 0$, we will have $Q(r)=\delta(r)$ leading to a saturating bound in the limit of $\avg{\Ss} \to 0$, as usual for uncertainty relations.

\begin{figure}
  \centering
  \includegraphics[width=\linewidth]{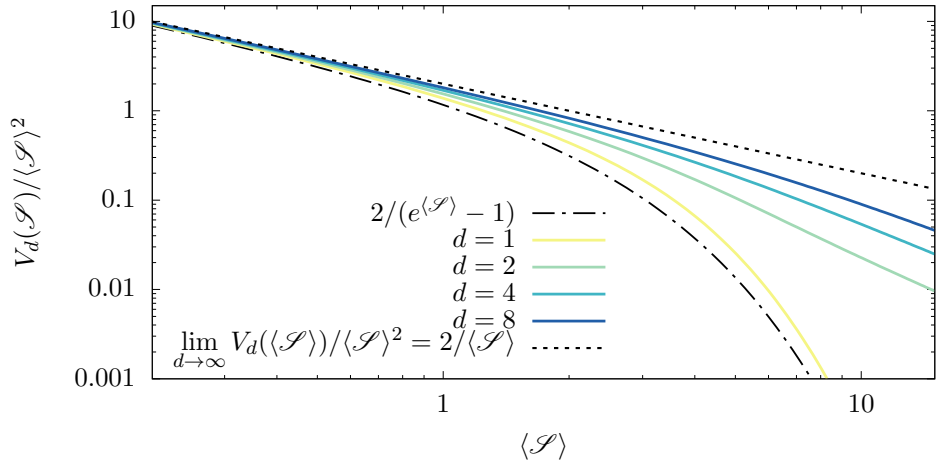}
  \caption[Caption ]{Plot of the symmetry-dependent lower bound~\eqref{eq:main} on the uncertainty ratio $ \textrm{Var} (\Ss)/\avg{\Ss}^2$ of the symmetry breaking order parameter as a function of the mean value of that order parameter.
If the broken symmetry were the time reversal symmetry, then the plot would be that of the TUR bounds versus the dissipation, in which case the symmetry-derived bounds are known: the exponential bound (dot-dashed) from Ref.~\cite{van2019uncertainty} and the $d=1$ bound (yellow) from Ref.~\cite{Timpanaro2019Thermodynamic}.
As the symmetry dimensionality increases, the bound tightens, reaching the original TUR (dotted) in the $d \to \infty$ limit.}
  \label{fig:bounds}
\end{figure}

Finally, we are able to inspect the isometric thermodynamic uncertainty bound, $V_d(\avg{\Ss})/\avg{\Ss}^2$ as a function of the mean symmetry breaking $\avg{\Ss} = \vect{F} \cdot \vect{\Op}$ and the dimensionality of the symmetry, $d$.
The plot in Fig.~\ref{fig:bounds} demonstrates that the isometric TUR improves with increasing $d$ since higher-dimensional symmetries impose more constraints on the distribution of the order parameter\footnote{The monotonic tightening follows from the fact that an order parmameter distribution obeying a $d$ dimensional symmetry must also obey a $d-1$ dimensional symmetry.}.
In the $d = 1$ limit, the rotation group is $\mathbb{Z}_2$, so the symmetry constraint on the order parameter is actually the exact same as that imposed by the discrete time-reversal symmetry.
Thus it should be no surprise that the results collapse onto the known FT bounds from Refs.~\cite{Timpanaro2019Thermodynamic,van2019uncertainty}:
\begin{equation}
    \label{eq:exponentialBounds}
    \textrm{Var} (\Ss) \geq V_1(\avg{\Ss})= \f{\avg{\Ss}^2}{\sinh^2\left(e_1^{-1}(\avg{\Ss}/2)\right)} \geq \f{2 \avg{\Ss}^2}{e^{\avg{\Ss}} -1}
\end{equation}
where $e_1^{-1}(x)$ is the inverse function of $x\tanh{x}$ from Eq.~\eqref{eq:er}.

Interestingly, the $d \to \infty$ limit also returns a known bound, in this case the original TUR.
To see this, we use the large-$d$ expansion (see also App.~\ref{app:convexity})
\begin{equation}
e_d^{-1}\left(\f{\avg{\Ss}}{2}\right) = \sqrt{d \f{\avg{\Ss}}{2}} + \frac{\avg{\Ss}^{3/2}}{4\sqrt{2d}}+\mathcal{O}\left(\frac{1}{d}\right).
\end{equation}
to compute the limiting behavior of $V_d(\avg{\Ss}) = 4c_d(e_d^{-1}(\avg{\Ss}/2) - \avg{\Ss}^2$.
Recalling, Eq.~\eqref{eq:cr}, we get
\begin{align}
\nonumber \lim_{d \to \infty} V_d(\avg{\Ss}) &= \lim_{d \to \infty} \left(\sqrt{d \f{\avg{\Ss}}{2}} + \frac{\avg{\Ss}^{3/2}}{4\sqrt{2d}}\right)^2 - (d-1) \f{\avg{\Ss}}{2} - \avg{\Ss}^2 + \mathcal{O}\left(\frac{1}{d}\right)\\
&= 2\avg{\Ss},
\end{align}
which leads to the uncertainty bound of $2 / \avg{\Ss}$.
The $d \to \infty$ limit is particularly simple because the minimum-variance order parameter distribution must become Gaussian in this limit, which is essentially the same reason the $O(N)$ model becomes solvable in infinite dimensions.
Gaussian current fluctuations also underpin the original TUR for Markov jump processes, though the Gaussian nature of the $d \to \infty$ limit emerges for a different reason.
In the former case Gaussians emerge from a linear response bound whereas in the latter case they come from the large dimensionality limit.

It is appealing mathematically that degrees of symmetry provide a natural interpolation between the FT TUR and the original entropy production TUR, but this connection also demonstrates that all of the symmetry-driven TURs are necessarily weaker than the original TUR.
If, for example, one is interested in studying the example of the introduction---heat currents through a slab between two thermal reservoirs---the isometric TUR would apply, but so would the stronger original TUR.
The principal applicability of isometric TURs is therefore confined to cases where the original TUR is known to break down.
Even in those cases, the inequalities weaken as the observation time increases, a phenomena similar to the one discussed for $d = 1$ in Ref.~\cite{Horowitz2019}.

\section{Application to Broken Symmetry of Equilibrium Statistical Mechanical Systems}
\label{sec:equilibrium}

\subsection{Mapping to Equilibrium Problems}
\label{sec:MappingTo Equilibrium}

Broken symmetries and isometric fluctuation theorems have direct analogs in equilibrium systems. As such, the bounds we have described can be mapped to equilibrium inequalities on fluctuations in the appropriate static order parameter\cite{lacoste2014isometric,lacoste2015fluctuation,guioth2016thermodynamic}.
To demonstrate the connection between nonequilibrium dynamical fluctuations and equilibrium static fluctuations, it is useful to observe how the Boltzmann distribution for configurational fluctuations yields expressions that mirror Secs.~\ref{sec:brokenTR} and~\ref{sec:brokenR}.
We focus our discussion around a 1d spin chain of length $N$ with periodic boundary conditions.
As is standard, the spins can interact with one or more neighbors with some potential $V$ as well as with an external field $\vect{B}$ such that the probability of spin configuration $\left\{\vect{\sigma}\right\}$ is
\begin{equation}
P_{\vect{B}}\left(\left\{\vect{\sigma}\right\}\right) = \frac{e^{- H\left(\left\{\vect{\sigma}\right\}\right)}}{Z_{\vect{B}}}
\label{eq:boltz}
\end{equation}
with Hamiltonian
\begin{equation}
H\left(\left\{\vect{\sigma}\right\}\right) = V\left(\left\{\vect{\sigma}\right\}\right) + \sum_{i = 1}^N \vect{B} \cdot \vect{\sigma}_i,
\end{equation}
inverse temperature $\beta=1$, and partition function $Z_{\vect{B}}$.
To match the notation of \sref{sec:Setup}, we have highlighted that the distribution over spins, $P_{\vect{B}}$, is a function of the external field $\vect{B}$.
In the absence of an external field and with a rotationally invariant $V$, the probability of a spin configuration $\left\{\vect{\sigma}\right\}$ is unaltered by a global rotation of all spins.
That is to say, for some rotated configuration $\left\{\vect{\sigma}'\right\} = \bm{R}\left(\left\{\vect{\sigma}\right\}\right)$ consisting of an identical rotation of each spin, $P_0\left(\left\{\vect{\sigma}\right\}\right) = P_0\left(\left\{\vect{\sigma}'\right\}\right)$.
The external field will break the rotational symmetry, resulting in a relation between $P_{\vect{B}}(\left\{\vect{\sigma}\right\})$ and $P_{\vect{B}}(\left\{\vect{\sigma}'\right\})$ that mirrors Eqs.~\eqref{eq:RN1} and~\eqref{eq:orderParameter}.
Because $P_{\vect{B}}$ is simply given by the Boltzmann distribution in Eq.~\eqref{eq:boltz}, we have
\begin{equation}
\ln \frac{P_{\vect{B}}\left(\left\{\vect{\sigma}\right\}\right)}{ P_{\vect{B}}\left(\left\{\vect{\sigma}'\right\}\right)} =   \, \vect{B} \cdot (\bm{R} \vect{M} -\vect{M} ),
\label{eq:equilRN}
\end{equation}
where $\vect{M}$ is the net magnetization 
\begin{equation}
\vect{M} = \sum_{i=1}^N \vect{\sigma}_i.
\end{equation}
Comparing Eqs.~\eqref{eq:orderParameter} and~\eqref{eq:equilRN}, the order parameter analogous to $\vect{\Op}$ is thus $-2 \vect{M}$, yielding the equilibrium isometric fluctuation theorem~\cite{lacoste2014isometric,lacoste2015fluctuation}:
\begin{equation}
\frac{P_{\vect{B}}(\vect{M})}{P_{\vect{B}}(\vect{M}')}=e^{ \vect{B}\cdot(\vect{M}'-\vect{M})}.
\end{equation}

Using the results from the previous sections, one can show that this implies a bound on the variance of $\vect{B}\cdot\vect{M}$:
\begin{equation}
\textrm{Var}(\vect{B}\cdot\vect{M})\geq c_d\left(e_d^{-1}\left(\left\langle \vect{B}\cdot\vect{M}\right\rangle\right)\right) - \left\langle \vect{B}\cdot\vect{M}\right\rangle^2,
\label{eq:varm}
\end{equation}
where $d$ is the dimension of the spin space.
For the nonequilibrium fluctuations, we focused on the variance of $\vect{F} \cdot \vect{\Op}$, a dissipative measure of the symmetry breaking.
Here, the observable $\vect{B} \cdot \vect{M}$ also measures symmetry breaking, but it carries the interpretation of the energy of coupling between the system and the external field.

\subsection{Example}
\label{sec:Example}

\begin{figure}[htp]
  \centering
  \includegraphics[width=\linewidth]{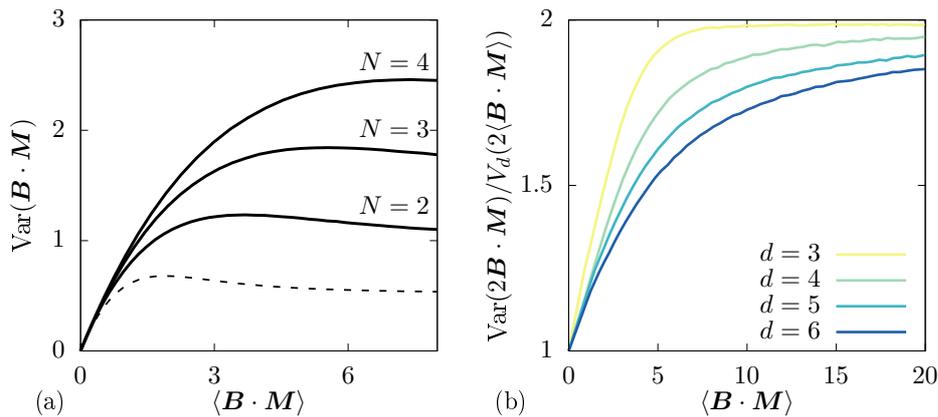}
  \caption{(a) Comparison of the bound (dashed line) and the variance (full line, obtained by Monte Carlo simulations) for a Heisenberg spin chain with spin dimension $d=2$ and chain length $N=2,3,$ and $4$. (b) Ratio between the variance and the bound for a chain of length $N=2$, for $d=3$ to $d=6$. 
For both figures, we set $J=0.1$ and converged the values by attempting $4\cdot 10^7$ Monte Carlo single-spin moves per data point.}
\label{fig:2}
\end{figure}

To evaluate the tightness of these uncertainty bounds, we turn to a Heisenberg spin chain model with periodic boundary conditions and nearest-neighbor interactions described by the Hamiltonian
\begin{equation}
  H = -J \sum_{i=1}^N \vect{\sigma}_i \cdot \vect{\sigma}_{i+1} + \sum_{i=1}^N \vect{B} \cdot \vect{\sigma}_i,
\end{equation}
with $\vect{\sigma}_i \in \mathbb{S}_d$ and coupling constant $J$ (not to be confused with the notation for currents at the beginning of the paper).
These spin chains provide a natural means to tune the system size $N$ and the dimensionality $d$.
In Fig.~\ref{fig:2}a we plot the bound on the variance in the energy of coupling between system and external field, Eq.~\eqref{eq:varm}.
This variance bound, plotted as a dashed line, notably does not depend on the system size $N$.
For comparison, we also plot the actual variance, computed from the Boltzmann distribution, Eq.~\eqref{eq:boltz}.
For $N=1$, the Boltzmann distribution for the spin corresponds to the Von Mises-Fisher distribution, Eq.~\eqref{eq:fisherVonMises} and therefore, the variance saturates the theoretical bound.
For $N>1$ variance exceeds the bound, but the tightness of the bound depends sensitively on the system size.
Specifically, larger systems exhibit larger fluctuations.
If the spins were all statistically independent ($J = 0$), the fluctuations in $\vect{B} \cdot \vect{M}$ would simply be proportional to $N$.
With coupled spins, the variance still grows with $N$ while the symmetry-derived bound does not.
Consequently, the symmetry-derived bound becomes insignificant in the large-system-size limit, analogous to the problem with long-time limits of FT bounds~\cite{Horowitz2019}.
Fig.~\ref{fig:2}a also illustrates that, independent of $N$, the bound is saturated for $\left\langle\vect{B} \cdot \vect{M}\right\rangle\lesssim k_{\rm B}T = 1$.
This saturation is a general prediction from linear-response theory \cite{Barato2015,macieszczak2018unified}.

As $d$ increases, the bound improves, though the large $N$ problem dominates unless the system size is very small.
Nevertheless, we illustrate the $d$-dependence of the bound using a tiny $N=2$ spin chain.
The ratio between the variance and the bound is pictured in Fig.~\ref{fig:2}b for increasing $d$, showing two principle features.
First, the variance of the $N=2$ spin chain cannot exceed twice the bound.
This restriction has a simple origin.
For large values of $\avg{\vect{B} \cdot \vect{M}}$, the spins becomes effectively independent, causing the ratio between the variance and the bounds to converge to $N$.
More interestingly, as $d$ increases, the ratio of variance to bound moves closer to unity.
Indeed, for higher dimensional spin spaces, a higher number of degrees of freedom are constrained by the isometric fluctuation theorem and therefore the bound is tighter.
In fact, we expect the $d\to \infty$ case would result in a saturated bound since the spin chain becomes the $O(N)$ model (see the discussion of the $d \to \infty$ limit in \sref{sec:Derivation}).

\section{Discussion and future directions}
\label{sec:Discussion}
In this article, we have studied the effects of broken symmetry on fluctuations. 
We have shown that the isometric fluctuation theorem allows the decomposition of an order parameter's probability distribution into an angular part subject to a fluctuation theorem and an independent radial part.  
This decomposition then leads to dimensionality-dependent lower bounds on the variance, that originate exclusively from the isometric fluctuation theorem.
We have studied those bounds both for nonequilibrium dynamical systems and equilibrium systems. 
The bounds are shown to be effective for small values of the order parameter due to linear response.
However, the order parameter will grow extensively when one considers either long trajectories or large system sizes, and the bounds become arbitrarily weak in either of those limits.

It is natural to wonder if extensive systems could be usefully bounded by splitting the system into effectively independent components, each of which could be bounded by the symmetry-derived bounds.
To do so, one might imagine cutting the trajectories (or systems) into small blocks with size equal to the correlation time (length) so that all blocks behave statistically independently.
Such a procedure would produce bounds which scale with system size in the same manner as the actual fluctuations scale, though we think it is nontrivial to make such a procedure rigorous.

Though they would suffer from the same extensive scaling problem, we anticipate additional symmetry-derived bounds that rely on fluctuation relations other than the isometric FT.
For example, observables satisfying the anisotropic fluctuation theorem \cite{VillavicencioSanchez2014Fluctuation,PerezEspigares2015Spatial} should have similar bounds that now require angular distributions over an ellipse rather than a sphere.
Such an extension is conceptually similar, though the computations would become more complicated.
It may also be possible to utilize additional discrete broken symmetries~\cite{Gaspard2012Flutuations,Gaspard2012Broken}, though the strategy could not be copied directly since the analog of $e_d(r)$ would not generally be invertible.
Whether that problem can be circumvented to give a general framework for calculating fluctuation constraints from (broken) symmetries remains an interesting question for future research.

\appendix
\section{Analytical properties of the Fisher-Von Mises distribution}
\label{app:convexity}

\subsection{Properties of Bessel functions}
\label{app:besselprops}
To derive the main properties of $e_d(r)$ and $c_d(r)$, we first state some well-known properties of the modified Bessel function of the first order, $I_\nu(r)$, then show their direct consequences for $e_d(r)$ and $c_d(r)$ \cite{abramowitz1965handbook,laforgia2010some,DLMF,sidi2011asymptotics}. 
Firstly, the recurrence relation,
\begin{equation}
    I_\nu(r)=I_{\nu-2}(r)-\frac{2(\nu-1)}{r}I_{\nu-1}(r),
\end{equation}
is equivalent to
\begin{equation}
    e_d(r)=\frac{r^2}{e_{d-2}(r)}-(d-2).\label{rec}
\end{equation}
The derivative of the Bessel function can be written as
\begin{equation}
    \frac{d}{dr}I_\nu(r)=I_{\nu-1}(r)-\frac{\nu}{r}I_\nu(r),
\end{equation}
which, after some algebra, leads to
\begin{equation}
    r\frac{d}{dr}e_d(r)=r^2-e_d(r)^2+(2-d)e_d(r).\label{dedx}
\end{equation}
Furthermore, the inequality
\begin{equation}
    1-\frac{2\nu}{r}\frac{I_\nu(r)}{I_{\nu-1}(r)}<\frac{I_\nu(r)^2}{I_{\nu-1}(r)^2},
\end{equation}
gives an inequality for $e_d(r)$
\begin{equation}
    r^2-de_d(r)-e_d(r)^2<0.\label{ineq}
\end{equation}
Another useful inequality is
\begin{equation}
     I_{\nu}(r)\leq I_{\nu-1}(r),
\end{equation}
or
\begin{equation}
    e_d(r)\leq r.\label{xbound}
\end{equation}
Finally, we mention a large-$\nu$  expansion of $I_\nu(r)$:
\begin{equation}
    \label{eq:largeorderAsymptI}
    I_\nu(r)\approx\frac{r^\nu}{2^\nu\Gamma(\nu+1)}\left(1+\frac{r^2}{4(\nu+1)}\right),
\end{equation}
where $\Gamma(r)$ is the Gamma-function. This implies
\begin{equation}
    e_d(r)\approx \frac{r^2}{d}-\frac{r^4}{d^2(d+2)},\label{lard}
\end{equation}
for large $d$.

\subsection{Invertibility of $e_d(r)$}
\label{sec:invertibility-er}

The invertibility of $e_d(r)$ on $[0,+\infty)$ simply follows from the positivity of its derivative over $[0,+\infty)$.
That derivative is simplified using the relations from~\ref{app:besselprops}:
\begin{eqnarray}
    r\frac{d}{dr}e_d(r)&=&r^2-e_d(r)^2+(2-d)e_d(r)\nonumber\\
    &=&r^2(e_{d-2}(r)^2+(d-2)e_{d-2}(r)-r^2\geq 0.\label{de}
\end{eqnarray}
The first equality follows from Eq.~(\ref{dedx}), the second equality follows from Eq.~(\ref{rec}), and the inequality follows from Eq.~(\ref{ineq}). 
This demonstration of positivity completes the proof for the invertibility of $e_d(r)$.

\subsection{Convexity of $c_d(e_d^{-1}(u))$}
\label{sec:convexity-ce-1x}

We shall now prove the convexity of $c_d(e^{-1}_d(u))$. 
First we note that
\begin{equation}
    c_d(e^{-1}_d(u))=\left(e^{-1}_d(u)\right)^2-(d-1)u.
\end{equation}
By taking two derivatives, we obtain
\begin{equation}
    \frac{d^2}{du^2}c_d(e^{-1}_d(u))=\frac{2}{\left(e_d'(e_d^{-1}(u))\right)^3}\left(e_d'(e_d^{-1}(u))-e_d^{-1}(u)e_d''(e_d^{-1}(u))\right),
\end{equation}
where $'$ denotes a derivative.
Therefore the function is convex if and only if,
\begin{equation}
    \frac{d}{dr}e_d(r)\geq r\frac{d^2}{dr^2}e_d(r),\label{tp}
\end{equation}
where $r = e^{-1}_d(u)$.
We prove this relation via an induction-like argument: first we show that Eq.~(\ref{tp}) holds for  large $d$ and subsequently, we will prove that if the relation holds for dimension $d+2$, it should also hold  for dimension $d-2$. The fact that  Eq.~(\ref{tp}) holds for  large $d$ immediately follows from Eq.~(\ref{lard}). Therefore, we should just prove that the relation holds for dimension $d-2$, given that it holds for $d+2$. To see this, we first start from Eq.~(\ref{rec}), which can be used to relate $e_{d-2}(r)$ to $e_{d+2}(r)$:
\begin{eqnarray}
    e_{d-2}(r)&=&\frac{r^2}{e_{d}(r)+d-2}\\
    &=&\frac{r^2(e_{d+2}(r)+d)}{r^2+(d-2)(e_{d+2}(r)+d)}.
\label{eq:relateds}
\end{eqnarray}
Our goal of the inductive step is to demonstrate that
\begin{equation}\frac{d}{dr}e_{d-2}(r)\geq r\frac{d^2}{dr^2}e_{d-2}(r),\end{equation}
but the relation between $d-2$ and $d+2$ in Eq.~\ref{eq:relateds} allows us to rexpress this target as a slightly complicated inequality involving $e_{d+2}(r)$ instead of $e_{d-2}(r)$:
\begin{eqnarray}
    &&\left((d-2)e_{d+2}(r)+d(d-2)+r^2\right)r^2\frac{d^2}{dr^2}e_{d+2}(r)-2(d-2)r^2\left(\frac{d}{dr}e_{d+2}(r)\right)^2\nonumber\\&&+\left(7(e_{d+2}(r)+d)(d-2)-r^2\right)r\frac{d}{dr}e_{d+2}(r)-\left(8d-16\right){(e_{d+2}(r)+d)^2}\leq 0.\label{tp2}
\end{eqnarray}
Using the inductive hypothesis,
\begin{equation}r^2\frac{d^2}{dr^2}e_{d+2}(r)\leq r\frac{d}{dr}e_{d+2},\end{equation}
we prove Eq.~\eqref{tp2}.
To do so, we recognize that Eq.~\eqref{xbound} requires the coefficient of the $e_{d+2}''(r)$ term in~\eqref{tp2} to be positive:
\begin{equation}(d-2)e_{d+2}(r)+d(d-2)+r^2\geq 0,\end{equation}
provided $d \geq 2$.
As a consequence, the left hand side of Eq.~\eqref{tp2} is upper bounded by
\begin{equation}
    -2(d-2)\left(r\frac{d}{dr}e_{d+2}(r)-2\left(e_{d+2}(r)+d\right)\right)^2,
\end{equation}
which is strictly negative.
Thus the $d-2$ concavity follows from $d+2$ concavity, completing the proof.

\end{document}